\pdfoutput = 1
\documentclass[twoside]{dis07}
\usepackage[latin1]{inputenc}
\usepackage{graphicx,color}
\usepackage{wrapfig,rotating}
\usepackage{amssymb,amsmath,array,bm}

\def\sket#1{\big|{#1}\big)}

\newbox\charbox
\newbox\slabox
\def\s#1{{      
        \setbox\charbox=\hbox{$#1$}
        \setbox\slabox=\hbox{$/$}
        \dimen\charbox=\ht\slabox
        \advance\dimen\charbox by -\dp\slabox
        \advance\dimen\charbox by -\ht\charbox
        \advance\dimen\charbox by \dp\charbox
        \divide\dimen\charbox by 2
        \raise-\dimen\charbox\hbox to \wd\charbox{\hss/\hss}
        \llap{$#1$}
}}

\begin{document}
\title{QCD and Monte Carlo generators}

\author{Zolt\'an Nagy
%
%
\vspace{.3cm}\\
%
Theory Division,
CERN
CH-1211 Geneva 23, Switzerland\\
E-mail: \tt{Zoltan.Nagy@cern.ch}
}

\maketitle

\begin{abstract}
In this talk I gave a brief summary of leading order, next-to-leading order and shower calculations. I discussed the main ideas and approximations of the shower algorithms and the related matching schemes. I tried to focus on QCD issues and open questions instead of making a inventory of the existing programs. 
\end{abstract}

\section{Fix order calculations}
\subsection{Born level calculations}

The simplest calculation what one can do is the Born level fix order calculation. This calculation involves the phase space integral of the tree level matrix element square and the jet measurement function. The structure of the cross section is 
\begin{equation}
\sigma[F_J] = 
\int_m d\varGamma^{(m)}(\{p\}_m)
|{\cal M}(\{p\}_m)|^2 F_J(\{p\}_m)\;\;,
\end{equation}
where $d\varGamma^{(m)}(\{p\}_m)$ is the phase space integral measure, ${\cal M}(\{p\}_m)$ represents the $m$-paraton tree level matrix element and $F_J(\{p\}_m)$ is the jet measurement function that defines the physical observable.

This calculation is relatively simple. The integral free from the infrared and ultraviolet singularities. The matrix element is basically a complicated expression but it can be generated in a automated way. Several implementations can be found in the literature, {\sc Alpgen}, {\sc Grace}, {\sc Helac}, {\sc Madgraph} and {\sc Sherpa}\cite{MEgenerator}.  

We can say that the tree level cross sections can predict the shape of the cross sections but in general they have several defects: i) Since it is the leading order term in the strong coupling expansion the result strongly depends on the unphysical renormalization and factorization scheme. ii) The exclusive physical quantities suffer on large logarithms. In the phase space regions where these logarithms are dominant the predictions are unreliable. iii) In the Born level calculations every jet is represented by a single parton, thus we don't have any information about the jet inner structure. iv) On the other hand in a real measurement, in the detector we can see hadrons and every jet consists many of them. 
We are not able to consider hadroniziation effects in the Born level calculations. 

\subsection{Next-to-leading order calculations}

We can increase the precision of our theory (QCD) prediction by calculating the next term in the perturbative expansion, the next-to-leading order correction (NLO). However this is just one order higher to the Born cross section but the complexity of the calculations increases enormously. We have to face to algebraic and analytic complexity.

The naive structure of the NLO calculation is 
\begin{equation}
\sigma_{\rm NLO} = 
\int_N d\sigma^B 
+\int_{N+1}d\sigma^R
+\int_{N}d\sigma^V\;\;.
\end{equation}
Here $\sigma^{B}$, $\sigma^{R}$ and $\sigma^{V}$ correspond to the Born, real and virtual contributions, respectively. This expression is well defined only in
$d=4-2\epsilon$ dimension because both the real and virtual terms are singular separately in $d=4$ dimension, but their sum is finite. Thus we cannot calculate them separately, first we have to regularize this integral. In the real part the singularities comes from the phase space integral from the regions where a gluon becomes soft of two partons become collinear and the integral over these degenerated phase space regions leads to contributions those are proportional to $1/\epsilon$ and $1/\epsilon^{2}$. The infrared singularity structure of the virtual contributions is exactly the same but with opposite sign, thus they cancel each other. To achieve this cancellation we have to reorganize our calculation in such a way that can be carried out in $d=4$ dimension
\begin{equation}
\sigma_{\rm NLO} = 
\int_N d\sigma^B 
+\int_{N+1}\big[d\sigma^R
-d\sigma^A\big]_{\epsilon=0}
+\int_{N}\big[d\sigma^V
+\int_1 d\sigma^A\big]_{\epsilon=0}\;\;.
\end{equation}
Here we subtracted the approximated version of real contribution and added it back in different form. In the second term $d\sigma^{A}$ cancels the singularities of $d\sigma^{R}$ and it is safe to perform the integral in $d=4$ dimension while in the third term the explicit singularities of $d\sigma^{V}$ are cancelled by $\int_{1}d\sigma^{A}$, where we performed the integral over the unresolved phase space analytically.  It is important that the approximated real contribution has universal structure. This term is based on the soft and collinear factorization property of the QCD matrix elements. A general subtraction scheme was defined by Catani and Seymour \cite{CS} and the extenion of this method for massive fermions is also available \cite{CSmassive}.  

The NLO calculation can be carried out but it hasn't been automated like the Born level calculations. The most complicated processes what we can calculate are $2 \to 3$ type \cite{nlomax5}. To go beyond this limit we have to find an efficient way to compute the virtual correction. Recently we have had some very promising developement on this area \cite{loopcalc}.

With the NLO corrections we can significantly reduce the dependence on the renormalization and factorization scales but in some cases it is not enough and the NNLO is also required. In these calculations one of the jet is represented by two partons. This can give some minimal information about the inner jet structure but is still very poor. The exclusive quantities are still suffers on large logarithms and we are still not able to consider hadronization effects.

\subsection{Next-to-next-to-leading order calculation}

For some processes and/or jet observables it is important to know the cross sections at next-to-next-toleading order level. In this cases the NLO $K$-factor usually large even larger than $2$ which means that the NLO correction doesn't reduce the scale dependences. Recently some simple but important processes have been calculated using sector decomposition method \cite{nnlosector} and there are some ongoing developments on defining a general scheme for NNLO calculations \cite{nnlosub}.

\section{Leading order parton shower}

The fix order calculations are systematically defined order by order 
and usually give good description well the data over the phase space where the large $p_{T}$ event are the dominant. In any order we still have to deal with the presence of the large logarithms and we cannot consider hadronization effect. 

\begin{figure}[t]
\center
\includegraphics[width=10cm]{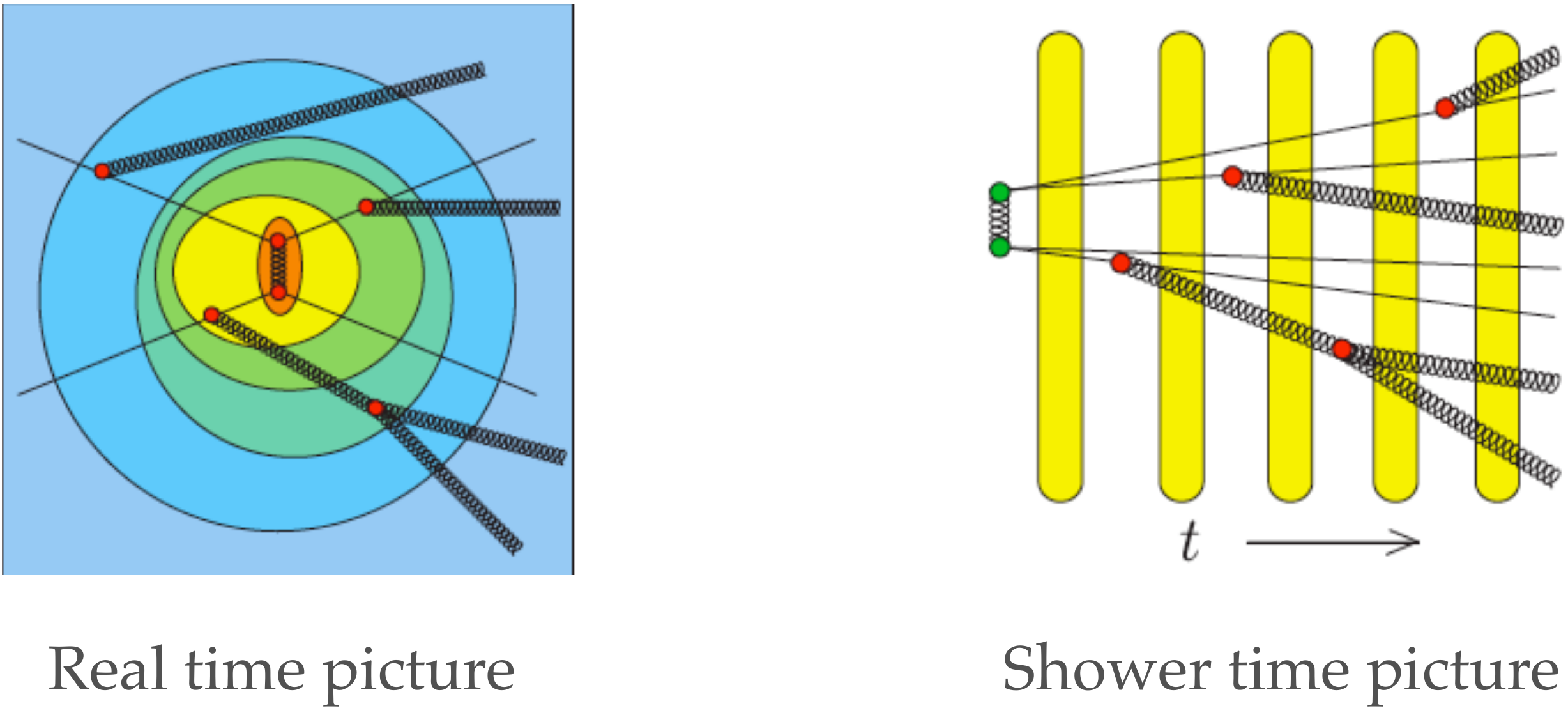}
\caption{\label{fig:shower}The left-hand picture depicts quark-quark scattering, with time proceeding from left to right. The hardest interactions are those toward the center of the picture. These are treated first in a parton shower Monte Carlo program. Thus in Monte Carlo time t, we start with the hard 
process and work toward softer interactions (with some of the particles moving backwards in real physical time), as depicted in the right hand picture. The rounded rectangles represent intervals of Monte Carlo time in which nothing happens.}
\end{figure}

There is an other way to calculate crass section in the perturbative framework, the parton shower calculations. Consider the parton shower picture of hadron-hadron scattering in which there is some sort of hard event, say jet production.
The parton shower description starts form hard scattering and proceeds forward to the softer scattering. In the final state the shower proceed forward in real time but for initial state parton the sowers proceeds backward in real time. This is depicted in Figure~\ref{fig:shower}.

\subsection{Shower evolution}

\begin{figure}[t]
\center
\includegraphics[width=10cm]{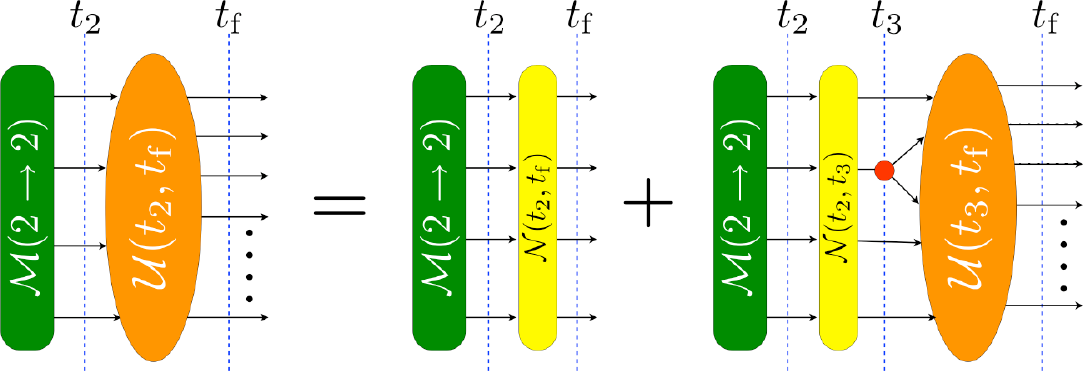}
\caption{\label{fig:evoleq} Evolution equation. The shower ({\em orange rectangle}) starts from the hard matrix elements ({\em green rounded rectangle}) and the partons are evolved to the final scale without splitting ({\em yellow rounded rectengle}) or with splitting at an intermediate time ({\em red circle}) and evolved to the finial scale with possible splittings. }
\end{figure}

The parton shower evolution can be represented by an evolution equation and it is the solution of the following integral equation
\begin{equation}
\label{eq:evol}
{\cal U}(t_{\rm f}, t_2)
\sket{{\cal M}_2} = 
{\cal N}(t_{\rm f}, t_2)
\sket{{\cal M}_2}
+
\int_{t_{2}}^{t_{\rm f}} dt_{3}\, {\cal U}(t_{\rm f}, t_3)
{\cal H}(t_3)
{\cal N}(t_3, t_2)
\sket{{\cal M}_2}\;\;.
\end{equation}
The shower evolution starts form the hard scattering and it is represented by the function $\sket{{\cal M}_2}$ that is a probability of a given partonic state in shower time $t_{2}$. Then  ${\cal U}(t_{\rm f}, t_2)$ is the probability function of having a particular partonic state in a later evolution time $t_{\rm f}$. The evolution operator is sum of two terms. The first term in Eq.~\eqref{eq:evol} represents parton evolution without splitting. The non splitting 
operator ${\cal N}(t_{\rm f}, t_2)$ that inserts Sudakov factors giving the 
probability that nothing happens between time $t_{2}$ and $t_{\rm f}$. The Sudakov is the exponentiated inclusive (summed over spin and color and integrated over the momenta of unresolved partons) splitting kernel. The second term in Eq.~\eqref{eq:evol} represent the splitting.  The partonic state is evolved without splitting to an intermediate time $t_{3}$ and splitting happens given by the splitting operator ${\cal H}(t_3)$ and the system is evolved with possible splitting from $t_{3}$ to $t_{\rm f}$. 
The splitting operator is based on the universal soft and collinear factorization property of the QCD matrix element. This evolution equation is depicted in Figure~\ref{fig:evoleq}.

\subsection{Splitting operator}

The splitting operator of the leading order (LO) shower is derived from the factorization property of the QCD matrix elements in the soft and collinear limits. This factorization property is universal. Let us start with collinear factorization. 

\begin{figure}[ht]
\center
\includegraphics[width=10cm]{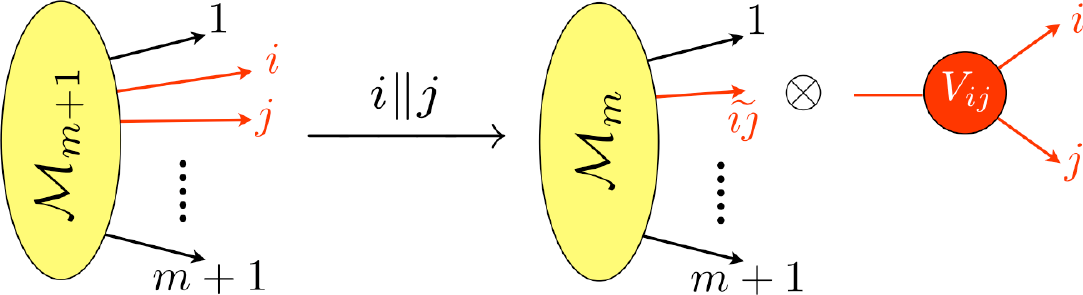}
\caption{\label{fig:collinear} Collinear limit. The universal singular part is represented by the red circle and it is based on the $1\to2$ matrix element at LO level and the yellow ovals represent the corresponding $m$ and $m+1$ parton matrix elements.}
\end{figure}

When two parton become collinear the $m+1$ tree level matrix element (and the matrix element square)  can be written as a convolution of $m$ parton hard matrix element and a universal singular factor in the spin space. This factorization is depicted in Figure~\ref{fig:collinear} at matrix element level. 

The collinear limit has some nice features. In the squared matrix element the singularity doesn't make color connection, it is completely factorized out. There are some spin correlation in the gluon splitting but this spin correlation is rather trivial. On the other hand one can always use spin averaged splitting functions but this is an additional approximation. Considering only the collinear emission our first ``candidate'' for the splitting kernel\footnote{For the shake of the simplicity I give a formal definition for the spin and color averaged splitting function which actually appears in the Sudakov exponent.}  would be
\begin{equation}
\label{collinearH}
\overline{\cal H}(t) = \sum_{l=1}^{m}\bm{T}^{2}_{l}\,V_{lm+1}(\hat p_{l},\hat p_{m+1}) 
\big[V_{lm+1}(\hat p_{l},\hat p_{m+1})\big]^{*}\;\;.
\end{equation}
Here $\bm{T}_{l}$ is the color charge operator of the mother parton and $V_{lm+1}$ is the vertex function. Every parton can split and after the splitting the daughter partons are labeled by $l$ and $m+1$. The momenta of the daughter partons are $\hat p_{l}$ and $\hat p_{m+1}$. With collinear splitting operator one can sum up the leading (double) logarithmic contributions properly.

\begin{figure}[t]
\center
\includegraphics[width=10cm]{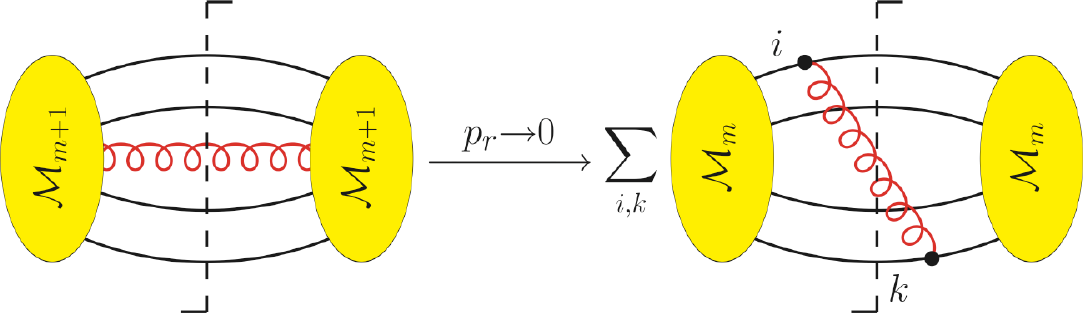}
\caption{\label{fig:soft} Soft factorization. The soft gluon is represented by the red wiggle line and the tree level matrix elements by the yellow ovals. The soft gluon can be emitted from any hard parton thus it makes color connections all possible way.}
\end{figure}

The matrix element has universal factorization property in the soft limit, when the energy of a final state gluon becomes zero. We cannot neglect soft gluon contributions since they produce next-to-leading logarithms when we integrate the matrix elements over the phase space. The structure of the soft gluon radiation is depicted in Figure~\ref{fig:soft}.
In this case we have non-trivial color structure because the soft gluon makes color connections all the possible way between the hard partons. Combining the soft factorization formulae with collinear one our splitting kernel is given by
\begin{equation}
\label{splitkernel}
\overline{\cal H}(t) = \sum_{l=1}^{m}\sum_{\substack{k=1\\k\neq l}}^{m}
\bm{T}_{l}\cdot\bm{T}_{k}\,
V_{lm+1}(\hat p_{l},\hat p_{m+1})
\big[
V_{km+1}(\hat p_{k},\hat p_{m+1})
- V_{lm+1}(\hat p_{l},\hat p_{m+1})\big]^{*}
\;\;.
\end{equation}
Here we used the color conservation, that is $\bm{T}_{l}^{2} = -\bm{T}_{l}\cdot\sum_{k\neq l}\bm{T}_{k}$.
One can show  that the expression under the square brackets vanishes in large angle limit when $\vartheta_{lk} \ll \vartheta_{lm+1},\,\vartheta_{km+1}$ (The $\vartheta_{ij}$ denotes the angle between momenta $p_{i}$ and $p_{j}$.). This effect is know as {\em color coherence}. Note, the color coherence breaks down with massive hard partons (quarks, SUSY particles). In this case we have wide angle soft radiations.

This is it, we defined a splitting kernel based on the soft and collinear approximation. With this we are able to sum up the leading and next-to-leading logarithms. We still have some freedom, for example the definition of splitting kernel away from the limits or the momentum mapping. The evolution (ordering) parameter can be basically any infrared sensitive variable such as the virtuality or the transverse momentum of the daughter partons. Note, if we want to consider spin and color correlations properly in the parton shower we {\em cannot} avoid negative weights. So far there is only one algorithm has been defined along this ideas \cite{NagySoper} but it hasn't been implemented yet. 
From the point of the implementation, the color interferences make some complications but one can impose some further approximations to simplify it.  

{\sc Herwing}\cite{herwig} and {\sc old Pythia}\cite{pythia} implement direct angular ordering\cite{angularorder}. Inserting the conditions for the emission angles into Eq.~\eqref{splitkernel} then  we have  
\begin{equation}
\label{angularorder1}
\overline{\cal H}(t) = -\frac12 \sum_{l=1}^{m}\sum_{\substack{k=1\\k\neq l}}^{m}
\bm{T}_{l}\cdot\bm{T}_{k}\,
\Big\{|V_{lm+1}|^{2}
\theta(\vartheta_{lm+1}< \vartheta_{lk})
+
|V_{km+1}|^{2}
\theta(\vartheta_{km+1} < \vartheta_{lk})
\Big\} + \cdots
\;\;,
\end{equation}
where the dots stand for the neglected terms those are finite in both soft and collinear limits. Note, these contributions are finite only after we perform the integral over the azimuthal angle of $\hat p_{m+1}$ about the direction $p_{l}$ and $p_{k}$. If we consider only those  phase space regions where emissions are ordered in angle then $\theta(\vartheta_{lm+1}< \vartheta_{lk}) = \theta(\vartheta_{km+1} < \vartheta_{lk})= 1$ and the color part of  Eq.~\eqref{angularorder1} becomes trivial and the approximated splitting kernel is identical to the collinear splitting kernel which is given in Eq.~\eqref{collinearH}. 

The other way to simplify the color structure is to expand the splitting kernel in powers of $1/N_{c}^{2}$, where $N_{c}$ is the number of the color states in fundamental representation. The gluon is a color $\bm{8}$, but in leading color approximation the gluon can be considered to be a $\bm{3}\otimes\bar{\bm{3}}$. The partons makes $\bm{3}\bar{\bm{3}}$ color dipoles with other partons but at leading color level the gluon never makes a color dipole with itself.
The color connection operator $\bm{T}_{l}\cdot\bm{T}_{k}$ becomes simple, it is non-zero if the partons $l$ and $k$ makes a $\bm{3}\bar{\bm{3}}$ color dipole and the approximated splitting operator is identical to the  collinear splitting kernel given in Eq.~\eqref{collinearH}. In this approximation it is important that the momentum mapping must be exact or based on dipole kinematics; if parton $l$ radiates a gluon then the recoiled parton must be the color connected one. 
{\sc Ariadne}\cite{ariadne} and the {\sc new Pythia}\cite{pythia} implement this approximation. There are some new developments \cite{ringbergZN} based on the leading color approximation and they implement this color dipole shower model.

The parton shower algorithms have been derived from perturbative QCD but we cannot consider them as theory predictions because they use rather nonsystematic approximations. The original idea was to consider and simulate higher order matrix element by using only soft and collinear factorization of the QCD matrix elements. This is a systematical approximation since the factorization properties of the matrix elements are held all order. At the end of this section  it is worthwhile to highlight the addition approximations and the limitation of the available parton shower implementations:
\begin{enumerate}
\item The current parton shower programs are still leading order calculations however they consider higher order contributions in an approximated way. Dependence on the unphysical scales is still strong.
\item The phase space is usually treated approximately. The angular ordered showers don't cover the phase space properly (``dead cone'') and some special treatment is required to to fill these regions.
\item The direct angular ordering or the leading color approximation neglect the color correlations. The color interferences could be significant in the case of non-global observables \cite{salamseymour}. Usually the spin correlations are also neglected. {\sc Herwig} considers spin correlations.
\item They are not defined systematically. The direct angular ordering is not defined or hard to define at higher order. Even the kinematics of the dipole shower model is inconsistent with the higher order. We have some freedom to define the splitting kernel and momentum mapping but the core algorithm should independent of the level of the calculation.
\item The only exact matrix element in the calculations is $2\to2$ like. If we want to calculate say $3, 4, 5,...$-jet cross section we should use $2\to 3, 4, 5,...$ LO or NLO matrix elements. In the next section I discuss the matching of shower to exact matrix elements.
\item More questions on non-perturbative effects: What is underlying event? How can we model it? How to consider quantum interferences in hadronization models?
\end{enumerate}

\section{Matching parton showers to fix order calculations}

\subsection{Born level matching}

\begin{figure}[t]
\center
\includegraphics[width=10cm]{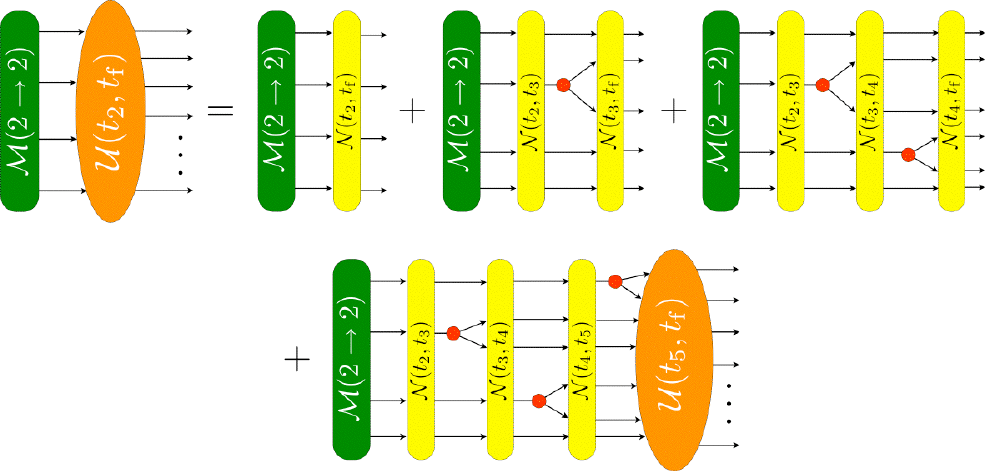}
\caption{\label{fig:shower2}Calculation of a shower starting with a $2 \to 2$ hard cross section (green rounded rectangle). The shower evolution operator has been iterated twice, so that the first term represents no splitting, the second term has one splitting, the third term has two splittings, and the final term contains contributions with three or more splittings.}
\end{figure}

\begin{figure}[ht]
\center
\includegraphics[width=10cm]{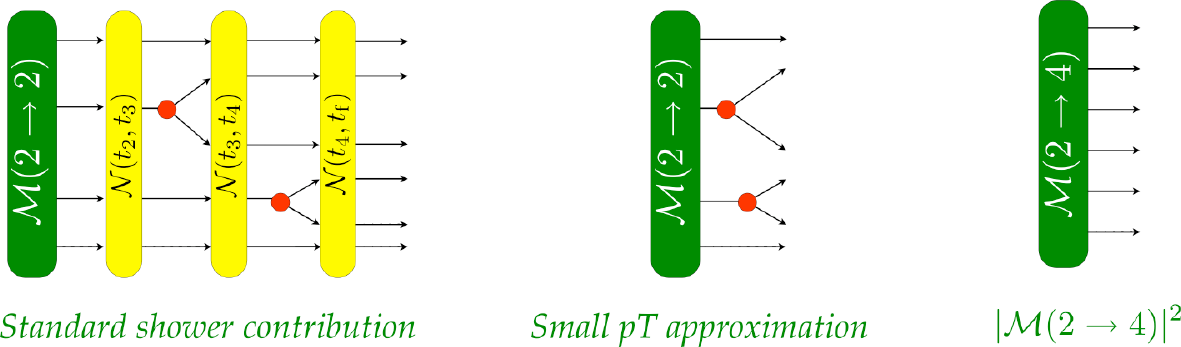}
\caption{\label{fig:showercontrib}The left-hand picture is the $2 \to 4$ cross section in shower approximation. The center picture is the shower approximation omitting the Sudakov factors. The right hand picture is the exact tree level $2 \to 4$ cross section. The cross section based on splitting functions (middle picture) is a collinear/soft approximation to this.
}
\end{figure}

The standard shower depicted in Figure~\ref{fig:shower2} has a deficiency, which is illustrated in Figure~\ref{fig:showercontrib}. 
The left-hand picture depicts a term contributing to the standard shower. In this term, there are Sudakov factors and $1 \to 2$ parton splitting functions. If we omit the Sudakov factors, we have the $1 \to 2$ parton splittings as depicted in the middle picture. These splittings are approximations based on the splitting angles being small or one of the daughter partons having small momentum. Thus the shower splitting probability with two splittings approximates the exact squared matrix element for $2 \to 4$ scattering. The approximation is good in parts of the final state phase space, but not in all of it. Thus one might want to replace the approximate squared matrix element of the middle picture with the exact squared matrix element of the right-hand picture. However, if we use the exact squared matrix element, we lack the Sudakov factors.

\begin{figure}[ht]
\center
\includegraphics[width=10cm]{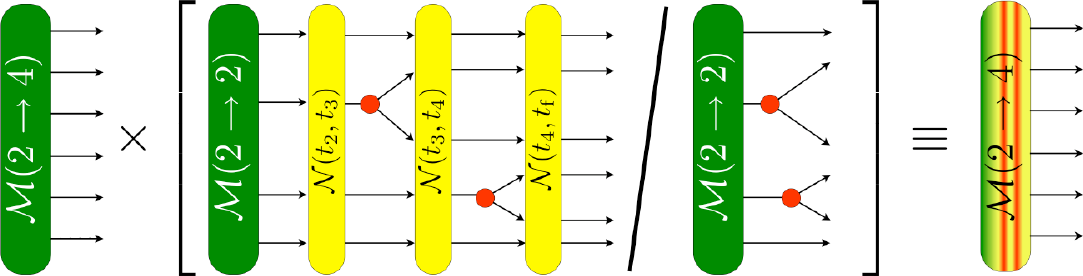}
\caption{\label{fig:sudakovreweight}An improved version of the $2 \to 4$ cross section. First we  generate the $4$-parton configuration according to the exact matrix element and take the shower approximation (with sudakov factors), divide by the approximate collinear squared matrix element, and multiply by the exact tree level squared matrix element. The graphical symbol on the right hand side represents this Sudakov reweighted cross section.}
\end{figure}

One can improve the approximation as illustrated in Figure~\ref{fig:sudakovreweight}. We reweight the exact squared matrix element by the ratio of the shower approximation with Sudakov factors to the shower approximation without Sudakov factors. The idea is to insert the Sudakov factors into the exact squared matrix element. This is the essential idea in the paper of Catani, Krauss, Kuhn, and Webber \cite{CKKW}. They use the $k_T$ jet algorithm to define the ratio needed to calculate the Sudakov reweighting factor.

\begin{figure}[ht]
\center
\includegraphics[width=10cm]{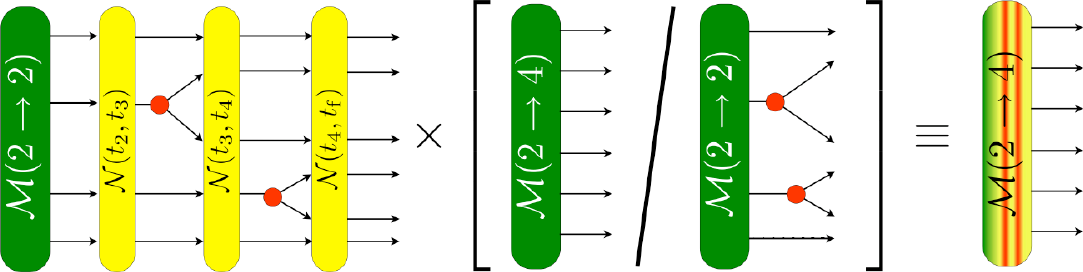}
\caption{\label{fig:mereweight}An improved version of the $2 \to 4$ cross section with matrix element reweighting factor. 
}
\end{figure}

There is another way to improve shower as illustrated in Figure~\ref{fig:mereweight}. First we generate the event according to the shower and then rewieght it by the ratio of the exact and approximated matrix element. The approximated matrix element is calculated  over a unique emission history that is determined by a jet algorithm. The original MLM algorithm \cite{MLM} uses the cone algorithm. The advantage of this method over the CKKW method is that the algorithm use the native Sudakov factors of the underlying parton shower. 

There is a further step in implementing this idea. CKKW divide the shower evolution into two stages, $0 < t < t_{\rm ini}$ and $t_{\rm ini} < t < t_{\rm f}$, where $t_{\rm ini}$ is a parameter that represents a moderate $P_T$ scale and $t_{\rm f}$ represents the very small $P_T$ scale at which showers stop and hadronization is simulated.

\begin{figure}[ht]
\center
\includegraphics[width=10.5cm]{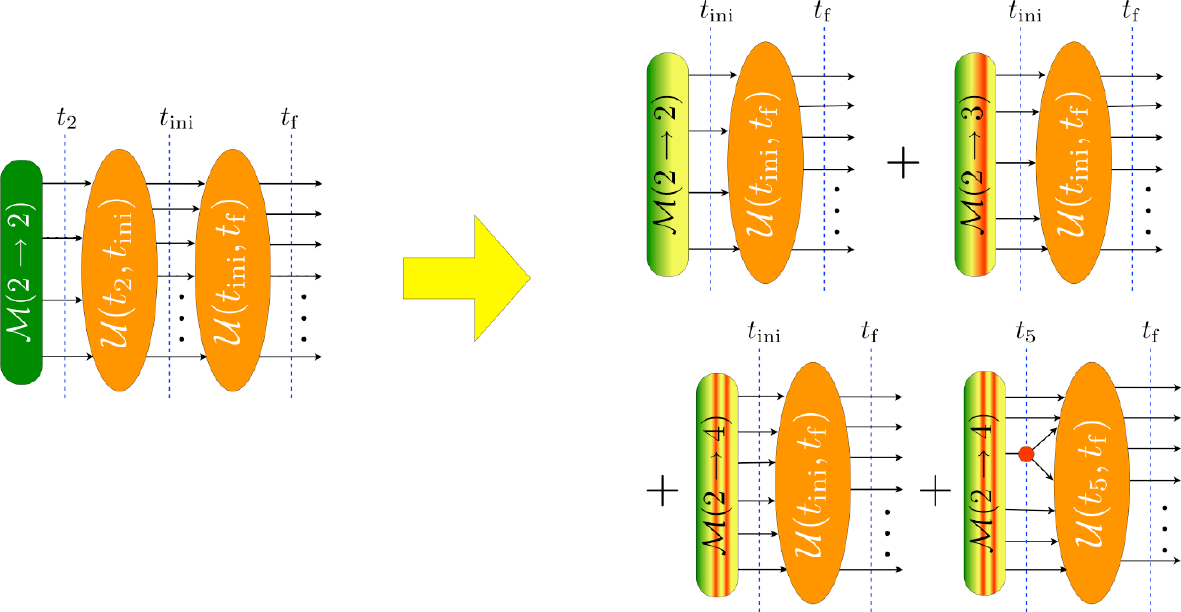}
\caption{\label{fig:ckkw}Shower with CKKW jet number matching. The calculation for $n$ jets at scale $t_{\rm ini}$ is based on the Sudakov reweighted tree level cross section for the production of $n$ partons.
}
\end{figure}

With this division, the Sudakov reweighting can be performed for the part of the shower at scale harder than $t_{\rm ini}$, as depicted in Figure~\ref{fig:ckkw}. The first term has no splittings at scale harder than $t_{\rm ini}$. In the second term there is one splitting, generated via the exact matrix element with a Sudakov correction as discussed above. In the next term there are two splittings. If we suppose that we do not have exact matrix elements for more than $2 \to 4$ partons, states at scale $t_{\rm ini}$ with more partons are generated with the ordinary parton shower. However, this contribution is suppressed by factors of $\alpha_s$. Evolution from $t_{\rm ini}$ to $t_{\rm f}$ is done via the ordinary shower algorithm.

To state the main idea of this jet number matching in a little different language, we can consider the cross section for an observable $F$. In the CKKW method, we break $\sigma[F]$ into a sum of contributions $\sigma_m[F]$ from final states with $m$ jets at resolution scale $t_{\rm ini}$. Then $\sigma_m[F]$ is evaluated using the exact tree level matrix element for $2 \to m$ parton scattering, supplemented by Sudakov reweighting and further supplemented by showering of the $m+2$ partons at scales softer than $t_{\rm ini}$. If $F$ is an infrared safe observable, this method gets $\sigma_m[F]$ correct to the leading perturbative order, $\alpha_s^m$. The method can be extended. The present authors have shown (at least for the case of electron-positron annihilation) how to get $\sigma_m[F]$ for an infrared safe observable correct to next-to-leading order, $\alpha_s^{m+1}$ \cite{NagySoperJHEP}. The required NLO adjustments are a little complicated, so I do not discuss them here.

\subsection{Next-to-leading order matching}

Matching parton shower with NLO fix order calculation is a very active field of parton shower developments. There are two basic approaches. First one is the {\sc MC@NLO} project \cite{mc@nlo}. The main idea here is to avoid the double counting by introducing extra counterterm which is extracted out from the underlying shower algorithm.  This method has been applied for several $2\to0+X$ where no colored object in the final state and some $2\to1+X$, $2\to2+X$ processes, where the QCD particles in the final state are heavy \cite{mc@nlo}.

The other approach was originally proposed by Kr\"amer and Soper \cite{kramersoper} and they implemented it for $e^{+}e^{-}\to 3$-jets. 
The idea is to include the first step of the shower in the NLO calculation and then start the parton shower from this configuration. Based on this concept some matching algorithm have been proposed but they are haven't been implemented \cite{ringbergZN,NagySoperJHEP,nason}. In the next I discuss in detail only the {\sc MC@NLO} approach because only this scheme has been implemented for LHC processes so far.

Let us start with the NLO cross section. After applying a subtraction scheme to remove the infrared singularities, we have
\begin{equation}
\sigma_{\rm NLO} =
\int_m\big[d\sigma^B + d\sigma^V+ d\sigma^C
+\int_1d\sigma^A\big] 
F_{\rm J}^{(m)}
+
\int_{m+1} \left[d\sigma^{\rm R}F_{\rm J}^{(m+1)} 
-d\sigma^{\rm A}F_{\rm J}^{(m)}\right]\;\;,
\end{equation} 
where $d\sigma^B$, $d\sigma^{\rm R}$, $d\sigma^V$, $d\sigma^C$ and $d\sigma^A$ are the Born, real, virtual contributions, collinear counterterm and subtraction term of the NLO scheme, respectively. The physical quantity is defined by the functions $F_{\rm J}^{(m)}$ and $F_{\rm J}^{(m+1)}$.

The naive way to add parton shower corrections is to replace the jet functions with the shower interface function. This approach is not good because it leads to double counting. It is easy to see, the shower that starts from the Born term generates higher order contributions those are already considered by NLO terms.  

To avoid double counting Frixione and Webber \cite{mc@nlo} organized the calculation in the following way:
\begin{equation}
\begin{split}
\sigma_{\rm MC} = {}&
\int_m\big[d\sigma^B + d\sigma^V+ d\sigma^C
+\int_1d\sigma^A\big] 
I_{\rm MC}^{(2\to m)}
\\
&+\int_{m+1} \left[d\sigma^{\rm R}_{m+1} 
- d\sigma^{\rm MC}_{m+1}\right]
I_{\rm MC}^{(2\to m+1)}
+ \int_{m+1} \left[
 d\sigma^{\rm MC}_{m+1} 
- d\sigma^{\rm A}_{m+1}\right]
I_{\rm MC}^{(2\to m)}\;\;.
\end{split}
\end{equation}   
Here the contribution $d\sigma^{\rm MC}_{m+1}$ is extracted from the underlying parton shower algorithm. The functions $I_{\rm MC}^{(2\to m)}$ and $I_{\rm MC}^{(2\to m+1)}$ are the interface functions to the shower. We have different choices for the $m$ and $m+1$ parton interface functions, thus we have
\begin{equation}
I_{\rm MC}^{(2\to m)} \sim {\cal U}(t_{\rm f}, t_{m})\qquad\text{and}\qquad
I_{\rm MC}^{(2\to m+1)}\sim{\cal U}(t_{\rm f}, t_{m+1}){\cal N}(t_{m+1}, t_{m})
\;\;.
\end{equation}
In the $m$-parton case we simply start the shower from the $m$-parton configuration while in the $m+1$ parton case first we insert some Sudakov factor representing the probability of nothing happens between the $m$-parton and $m+1$ parton states and starts the shower from the $m+1$ parton configuration.   

There are some limitation of the MC@NLO approach: i) It is worked out for {\sc Herwig}. One has to redo the Monte Carlo subtraction scheme if we want to match say {\sc Pythia} to NLO computations. ii) Matching procedure is defined only for simple processes like $2\to 0+X,1+X,2$. iii) The double counting problem is not fully solved but it is probably numerically invisible because of the strong Sudakov suppression. The problem related to the soft singularities and it appears only in the $2\to2$ like processes where the color structure is not trivial. 

\section{Conclusions}

Parton shower event generators have proved to be an essential tool for particle physics. These computer programs perform calculations of cross sections according to an approximation to the standard model or some of its possible extensions. Because of the great success of these programs, it is worthwhile to investigate possible improvements.

In a typical parton shower event generator, the physics is modeled as a process in classical statistical mechanics. Some number of partons are produced in a hard interaction. Then each parton has a chance to split into two partons, with the probability to split determined from an approximation to the theory. Parton splitting continues in this probabilistic style until a complete parton shower has developed.

The underlying approximation is the factorization of amplitudes in the soft or collinear limits. However, further approximations are usually added:
i) The interference between a diagram in which a soft gluon is emitted from one hard parton and a diagram in which the same soft gluon is emitted from another hard parton is treated in an approximate way,  with the ``angular ordering'' approximation.
ii) Color is treated in an approximate way, valid when $1/N_{\rm c}^2 \to 0$ where $N_{\rm c} = 3$ is the number of colors.
iii) Parton spin is treated in an approximate way. According to the full quantum amplitudes, when a parton splits, the angular distribution of the daughter partons depends on the mother parton spin and even on the interference between different mother-parton spin states. This dependence is typically ignored.
With the use of these further approximations, one can get to a formalism in which the shower develops according to classical statistical mechanics with a certain evolution operator.

I think the way to improve the parton showers is to formulate it based on the factorization of amplitudes in the soft or collinear limits in which one does not make the additional approximations enumerated above. For this, one would have to use quantum statistical mechanics instead of classical statistical mechanics. 

On the other hand the parton shower algorithm should cooperate with exact LO and NLO matrix elements. Currently we have some very promising tools such as CKKW, MLM and MC@NLO matching schemes. The CKKW and MLM matching procedures patch the ``hole'' between the Born level fix order and the shower calculations while the MC@NLO and other NLO matching schemes do the same between the shower and fix order NLO calculations.
If we want more precise tools we need more advanced framework. We need a general LO shower framework that naturally includes the LO and NLO calculation. Or phrase it differently, we should reformulate the LO and NLO calculation to make the shower part of them.

\section{Acknowledgment}

I am greatful to the organizers of the DIS 2007 workshop for their invitation 
as well as for providing a pleasant atmosphere during the meeting.
This work was supported by the Hungarian Scientific Research Fund grants OTKA T-60432.

\begin{footnotesize}

\end{footnotesize}

\begin{thebibliography}{99}
\bibitem{url} Slides: \\ 
\verb$http://indico.cern.ch/contributionDisplay.py?contribId=6&sessionId=2&confId=9499$
\bibitem{MEgenerator}
  M.~L.~Mangano, M.~Moretti, F.~Piccinini, R.~Pittau and A.~D.~Polosa,
  JHEP {\bf 0307} (2003) 001
  [arXiv:hep-ph/0206293]; 
  G.~Belanger, F.~Boudjema, J.~Fujimoto, T.~Ishikawa, T.~Kaneko, K.~Kato and 
  Y.~Shimizu,
  Phys.\ Rept.\  {\bf 430} (2006) 117
  [arXiv:hep-ph/0308080];
  A.~Kanaki and C.~G.~Papadopoulos,
  Comput.\ Phys.\ Commun.\  {\bf 132} (2000) 306
  [arXiv:hep-ph/0002082];
  F.~Maltoni and T.~Stelzer,
  JHEP {\bf 0302} (2003) 027
  [arXiv:hep-ph/0208156];
  T.~Gleisberg, S.~Hoche, F.~Krauss, A.~Schalicke, S.~Schumann and J.~C.~Winter,
  JHEP {\bf 0402} (2004) 056
  [arXiv:hep-ph/0311263].
\bibitem{CS}
  S.~Catani and M.~H.~Seymour,
  Nucl.\ Phys.\  B {\bf 485} (1997) 291
  [Erratum-ibid.\  B {\bf 510} (1998) 503]
  [arXiv:hep-ph/9605323]. 
\bibitem{CSmassive}
  S.~Catani, S.~Dittmaier, M.~H.~Seymour and Z.~Trocsanyi,
  Nucl.\ Phys.\  B {\bf 627} (2002) 189
  [arXiv:hep-ph/0201036].   
\bibitem{nlomax5}
  Z.~Nagy,
  Phys.\ Rev.\  D {\bf 68} (2003) 094002
  [arXiv:hep-ph/0307268];
    Z.~Nagy and Z.~Trocsanyi,
  Phys.\ Lett.\  B {\bf 634} (2006) 498
  [arXiv:hep-ph/0511328];
  J.~M.~Campbell, R.~Keith Ellis and G.~Zanderighi,
  JHEP {\bf 0610} (2006) 028
  [arXiv:hep-ph/0608194];
  J.~Campbell, R.~K.~Ellis and D.~L.~Rainwater,
  Phys.\ Rev.\  D {\bf 68} (2003) 094021
  [arXiv:hep-ph/0308195].
\bibitem{loopcalc}
  G.~Ossola, C.~G.~Papadopoulos and R.~Pittau,
  Nucl.\ Phys.\  B {\bf 763} (2007) 147
  [arXiv:hep-ph/0609007];
  T.~Binoth, J.~P.~Guillet and G.~Heinrich,
  JHEP {\bf 0702} (2007) 013
  [arXiv:hep-ph/0609054];
  Z.~Nagy and D.~E.~Soper,
  JHEP {\bf 0309} (2003) 055
  [arXiv:hep-ph/0308127];
  C.~F.~Berger, Z.~Bern, L.~J.~Dixon, D.~Forde and D.~A.~Kosower,
  Phys.\ Rev.\  D {\bf 74} (2006) 036009
  [arXiv:hep-ph/0604195]; 
  Z.~Bern, L.~J.~Dixon and D.~A.~Kosower,
  Phys.\ Rev.\  D {\bf 71} (2005) 105013
  [arXiv:hep-th/0501240];
  C.~Anastasiou, R.~Britto, B.~Feng, Z.~Kunszt and P.~Mastrolia,
  JHEP {\bf 0703} (2007) 111
  [arXiv:hep-ph/0612277];
  R.~Britto, B.~Feng and P.~Mastrolia,
  Phys.\ Rev.\  D {\bf 73} (2006) 105004
  [arXiv:hep-ph/0602178];
  T.~Binoth, G.~Heinrich, T.~Gehrmann and P.~Mastrolia,
  Phys.\ Lett.\  B {\bf 649} (2007) 422
  [arXiv:hep-ph/0703311]
\bibitem{nnlosector}
  G.~Heinrich,
  Nucl.\ Phys.\ Proc.\ Suppl.\  {\bf 116} (2003) 368
  [arXiv:hep-ph/0211144];
    T.~Binoth and G.~Heinrich,
  Nucl.\ Phys.\  B {\bf 693} (2004) 134
  [arXiv:hep-ph/0402265];
  C.~Anastasiou, K.~Melnikov and F.~Petriello,
  Phys.\ Rev.\ Lett.\  {\bf 93} (2004) 262002
  [arXiv:hep-ph/0409088].
\bibitem{nnlosub}
  A.~Gehrmann-De Ridder, T.~Gehrmann, E.~W.~N.~Glover and G.~Heinrich,
  Nucl.\ Phys.\ Proc.\ Suppl.\  {\bf 160} (2006) 190
  [arXiv:hep-ph/0607042];
  A.~Gehrmann-De Ridder, T.~Gehrmann and E.~W.~N.~Glover,
  JHEP {\bf 0509} (2005) 056
  [arXiv:hep-ph/0505111];
  G.~Somogyi, Z.~Trocsanyi and V.~Del Duca,
  JHEP {\bf 0701} (2007) 070
  [arXiv:hep-ph/0609042];
  G.~Somogyi and Z.~Trocsanyi,
  JHEP {\bf 0701} (2007) 052
  [arXiv:hep-ph/0609043];
  S.~Catani and M.~Grazzini,
  arXiv:hep-ph/0703012;
  S.~Weinzierl,
  Phys.\ Rev.\  D {\bf 74} (2006) 014020
  [arXiv:hep-ph/0606008].
\bibitem{NagySoper}
  Z.~Nagy and D.~E.~Soper,
  arXiv:0706.0017 [hep-ph].
\bibitem{herwig}
  G.~Marchesini, B.~R.~Webber, G.~Abbiendi, I.~G.~Knowles, 
  M.~H.~Seymour and L.~Stanco,
  Comput.\ Phys.\ Commun.\  {\bf 67} (1992) 465 ;
  S.~Gieseke {\it et al.},
  [arXiv:hep-ph/0609306].
  
\bibitem{pythia}
  T.~Sj\"ostrand,
  Comput.\ Phys.\ Commun.\  {\bf 82} (1994) 74; 
  T.~Sj\"ostrand, S.~Mrenna and P.~Skands,
  JHEP {\bf 0605} (2006) 026 
  [arXiv:hep-ph/0603175].
\bibitem{angularorder}
  G.~Marchesini and B.~R.~Webber,
  Nucl.\ Phys.\ B {\bf 238} (1984) 1;
  R.~K.~Ellis, G.~Marchesini and B.~R.~Webber,
  Nucl.\ Phys.\ B {\bf 286} (1987) 643 
  [Erratum-ibid.\ B {\bf 294} (1987) 1180].
\bibitem{ariadne}
  L.~L\"onnblad,
  Comput.\ Phys.\ Commun.\  {\bf 71} (1992) 15.
\bibitem{ringbergZN}
  Z.~Nagy and D.~E.~Soper,
  arXiv:hep-ph/0601021.
\bibitem{salamseymour}
  J.~R.~Forshaw, A.~Kyrieleis and M.~H.~Seymour,
  JHEP {\bf 0608} (2006) 059
  [arXiv:hep-ph/0604094];
  M.~Dasgupta and G.~P.~Salam,
  Phys.\ Lett.\  B {\bf 512} (2001) 323
  [arXiv:hep-ph/0104277].
\bibitem{CKKW}
  S.~Catani, F.~Krauss, R.~Kuhn and B.~R.~Webber,
  JHEP {\bf 0111} (2001) 063
  [arXiv:hep-ph/0109231]; 
  L.~L\"onnblad,
  JHEP {\bf 0205} (2002) 046
  [arXiv:hep-ph/0112284];
  N.~Lavesson and L.~L\"onnblad,
  JHEP {\bf 0507} (2005) 054
  [arXiv:hep-ph/0503293];
\bibitem{MLM}
  M.~L.~Mangano, M.~Moretti, F.~Piccinini and 
  M.~Treccani,
  JHEP {\bf 0701} (2007) 013
  [arXiv:hep-ph/0611129].
\bibitem{NagySoperJHEP}
  Z.~Nagy and D.~E.~Soper,
  JHEP {\bf 0510} (2005) 024
  [arXiv:hep-ph/0503053].
\bibitem{mc@nlo}
  S.~Frixione and B.~R.~Webber,
  JHEP {\bf 0206} (2002) 029
  [arXiv:hep-ph/0204244];
  S.~Frixione, P.~Nason and B.~R.~Webber,
  JHEP {\bf 0308} (2003) 007
  [arXiv:hep-ph/0305252];
  S.~Frixione and B.~R.~Webber,
  arXiv:hep-ph/0612272.
\bibitem{kramersoper}
  M.~Kr\"amer and D.~E.~Soper,
  Phys.\ Rev.\  D {\bf 69} (2004) 054019
  [arXiv:hep-ph/0306222];
  D.~E.~Soper,
  Phys.\ Rev.\  D {\bf 69} (2004) 054020
  [arXiv:hep-ph/0306268];
  M.~Kr\"amer, S.~Mrenna and D.~E.~Soper,
  Phys.\ Rev.\  D {\bf 73} (2006) 014022 
  [arXiv:hep-ph/0509127].
\bibitem{nason}
  P.~Nason,
  JHEP {\bf 0411} (2004) 040
  [arXiv:hep-ph/0409146].











\end{thebibliography}
\end{document}